%% file: acl_latex.tex
\pdfoutput=1

\documentclass[11pt]{article}

\usepackage{acl}

\usepackage{times}
\usepackage{latexsym}

\usepackage[T1]{fontenc}

\usepackage[utf8]{inputenc}
\usepackage{graphicx}
\usepackage[calcwidth = .9\linewidth, font = small, labelfont = bf]{caption}
\usepackage{enumitem}
\usepackage{booktabs}
\usepackage{multirow}
\usepackage{tikz}
\usepackage{array}
\usepackage{soul}
\usepackage{array}

\usepackage{tablefootnote}
\usepackage[leftmargin=1em]{quoting}

\newcommand*\circled[1]{\tikz[baseline=(char.base)]{
            \node[shape=circle,draw,inner sep=1pt] (char) {#1};}}

\usepackage{microtype}

\input{macros}

\title{\large Augmenting Scientific Creativity with Retrieval across Knowledge Domains}

\author{Hyeonsu B. Kang\textsuperscript{1*}\quad 
        Sheshera Mysore\textsuperscript{2*}\quad 
        Kevin Huang\textsuperscript{3}\thanks{*Equal contribution.}\quad
        {\bf Haw-Shiuan Chang}\textsuperscript{2}\quad\\
        {\bf Thorben Prein}\textsuperscript{3}\quad
        {\bf Andrew McCallum}\textsuperscript{2}\quad
        {\bf Aniket Kittur}\textsuperscript{1}\quad
        {\bf Elsa Olivetti}\textsuperscript{3}\quad\\
        \textsuperscript{1}Human-Computer Interaction Institute, \textsc{\small CMU, PA, USA}\\
        \textsuperscript{2}Manning College of Information and Computer Sciences,  \textsc{\small UMass Amherst, MA, USA}\\ 
        \textsuperscript{3}Department of Materials Science and Engineering,  \textsc{\small MIT, MA, USA}\\
  \texttt{\{hyeonsuk,nkittur\}@cs.cmu.edu}\\\texttt{\{smysore, hschang, mccallum\}@cs.umass.edu}\\\texttt{\{kjhuang, prein, elsao\}@mit.edu}}

\begin{document}
\maketitle
\begin{abstract}
\input{src0_abstract}

\end{abstract}

\input{src0_intro}

\input{src1_system}

\input{src2_study}

\input{src3_case_study_findings}

\input{src3_implications}

\input{src4_conclusion}

\bibliography{anthology,zz_bib}
\bibliographystyle{acl_natbib}

\appendix

\label{sec:appendix}

\input{src5_appendix}

\end{document}

%% file: macros.tex
\newcommand{\xhdr}[1]{\vspace{1mm}\noindent{{\bf #1.}}} 

\newcolumntype{P}[1]{>{\centering\arraybackslash}p{#1}}
\definecolor{lgrey}{gray}{0.9}
\newcommand*\cirnum[1]{\raisebox{.5pt}{\textcircled{\raisebox{-.9pt} {#1}}}}

%% file: src0_abstract.tex
\vspace{-.5em} Exposure to ideas in domains outside a scientist's own may benefit her in reformulating existing research problems in novel ways and discovering new application domains for existing solution ideas. While improved performance in scholarly search engines can help scientists efficiently identify relevant advances in domains they may already be familiar with, it may fall short of helping them explore diverse ideas \textit{outside} such domains. In this paper we explore the design of systems aimed at augmenting the end-user ability in cross-domain exploration with flexible query specification. To this end, we develop an exploratory search system in which end-users can select a portion of text core to their interest from a paper abstract and retrieve papers that have a high similarity to the user-selected core aspect but differ in terms of domains. Furthermore, end-users can `zoom in' to specific domain clusters to retrieve more papers from them and understand nuanced differences within the clusters. Our case studies with scientists uncover opportunities and design implications for systems aimed at facilitating cross-domain exploration and inspiration.

%% file: src0_intro.tex
\section{Introduction} \vspace{-.5em}
Analogies have been a central mechanism for cross-boundary inspirations and innovation throughout the history of science and technology~\cite{gentnerAnalogyCreativityWorks1997,oppenheimerAnalogyScience1956}. For example, analogical reasoning helped the Greek philosopher Chrysippus (c. \textsc{240 b.c.}) model sound waves from observations of water waves, two domains that did not interact prior to this insight. However, as knowledge areas deepen in specialization, they interact less with each other~\cite{swanson1996undiscovered, chu2021slowed} and analogical reasoning across domains becomes increasingly challenging. 
Yet, a systematic review of existing literature suggests that enabling connections that \textit{jump} between scientific domains may bring significant innovations and scientific progress \cite{rzhetsky2015choosing}. As such, developing a computational means to augmenting human intelligence with analogical creativity remains a holy-grail challenge in both cognitive science and AI research~\cite{mitchell1993analogy,hesse1965models,kittur2019scaling}.

To this end, recent approaches for retrieving analogically similar ideas have demonstrated promising results~\cite{kang2022augmenting,hope_kdd17,hope2021scaling,chan2018solvent}. The central idea of these approaches focuses on retrieving analogs similar in one aspect of a pre-determined schema (e.g., `problem solved') while differing on the other (e.g., `method used'), exposure to which has shown to unlock scientists' creativity to break out of fixation and mechanistic conventions~\cite{kang2022augmenting}.

While promising, there are remaining technical and design challenges that prohibit realization of systems that support exploration across boundaries of knowledge domains. The first challenge is developing mechanisms that allow end-users to explore diverse domains and find ones they want to jump to find analogous retrievals.
Prior work in cognitive psychology research on human creativity~\cite{chanBestDesignIdeas2015,goncalvesInspirationPeakExploring2013} has noted a sweet spot of distance for retrieval, with effective results being neither too different from the query, which may cause early rejection from searchers, nor too similar, which may be insufficient for breaking out of conventions and fixation~\cite{jansson1991design}. This notion of effective distance is a departure from the commonly used notions of relevance in the literature of IR and NLP that rely on topical overlap~\cite{borlund2003concept} and development of methods often optimized for maximal content-similarity~\cite{carterette2011system}. A notable exception is the prior research in search diversification (see ~\citet{santos2015diversesurvey}), which contributes mechanisms for \emph{optimizing} the novelty and coverage of search results given ambiguous queries. However, they often lack the controllability for end-users who want to specify the kinds of novelty they are interested in and the scope of coverage. 

The second challenge is the lack of labeled data for scaling a schematic representation of documents. Barriers to domain knowledge make accurate labeling of research papers using a pre-determined schema across diverse domains prohibitively costly. While~\citet{chan2018solvent} demonstrated a potential pipeline for crowdsourcing the labels with non-expert crowdworkers, a subsequent paper in which a supervised model was trained based on the crowd-labeled data showed limitations on the model accuracy and downstream retrieval usefulness~\cite{kang2022augmenting}. Furthermore, emerging scientific domains present an additional challenge with updating data.

Finally, studying how cross-domain explorations occur in authentic use scenarios requires a) development of an interactive search system with a usable performance and b) evaluating it with scientists seeking inspirations on real-world research problems. Insights from the study also contribute to the growing literature that studies open questions around how informational and exploration needs in creative work change over time and during the exploration~\cite{li2022analyzing}.

In this work we develop an interactive prototype system which we use to probe these challenges and uncover design implications for future analogical search systems aimed at supporting retrieval across knowledge domains\footnote{ \url{https://github.com/olivettigroup/cross-domain-exploration}}. Our system leverages the recently introduced \textit{faceted Query-by-Example (QbE)} paradigm to increase the controllability of search novelty and scope, by allowing searchers to query using a paper abstract {and} a sentence\footnote{Future systems may extend this to support sub-sentence spans or multiple-sentence queries.} in the abstract that indicates personally interesting core aspects of the paper~\cite{MysoreCSFCUBE2021}. To find matching abstracts based on the query aspect, our system first indexes documents using each of its sentences and retrieves similar sentences using a strong sentence encoding model~\cite{MPNetBaseV2} at runtime. To go beyond the high-demand for labeled data in a challenging domain, our approach relaxes the mechanism for retrieval from relying on a pre-determined schematic representation of abstracts to using a readily available, intuitive sentence-level representation.
To support an increased diversity of matching over prior work that mapped each abstract into a single `best' schematic representation, we enable users to select any sentence in an abstract as interesting for querying. The system then retrieves papers similar to the selected aspect but different in terms of their domains. Domain clusters in our system were constructed using a pre-trained scientific document representation model trained on citation data~\cite{cohan2020specter} capturing a global domain knowledge structure, and also do not require explicit labeled data in the process.

In order to center authentic user needs in the development of our prototype system, we adopted an iterative design process and sought feedback from a group of materials science researchers studying climate change mitigation strategies. Due to the complex systems nature of climate change, development of mitigation strategies such as clean energy technologies and more sustainable building materials {requires} a cross-disciplinary approach~\cite{xu2016interdisciplinary} and related closely to the needs we set out to support in this research. We conducted case studies with our system to gain deeper insights into whether the system retrieves results that represent meaningful `jumps' between knowledge domains, whether scientists can make sense of them, and the design challenges for future exploratory search systems that facilitate cross-boundary inspirations.

%% file: src1_system.tex
\section{System Description}
\vspace{-.5em}
\label{sec-system-descr}
\begin{figure}[t]
    \centering
    \includegraphics[width=.9\linewidth]{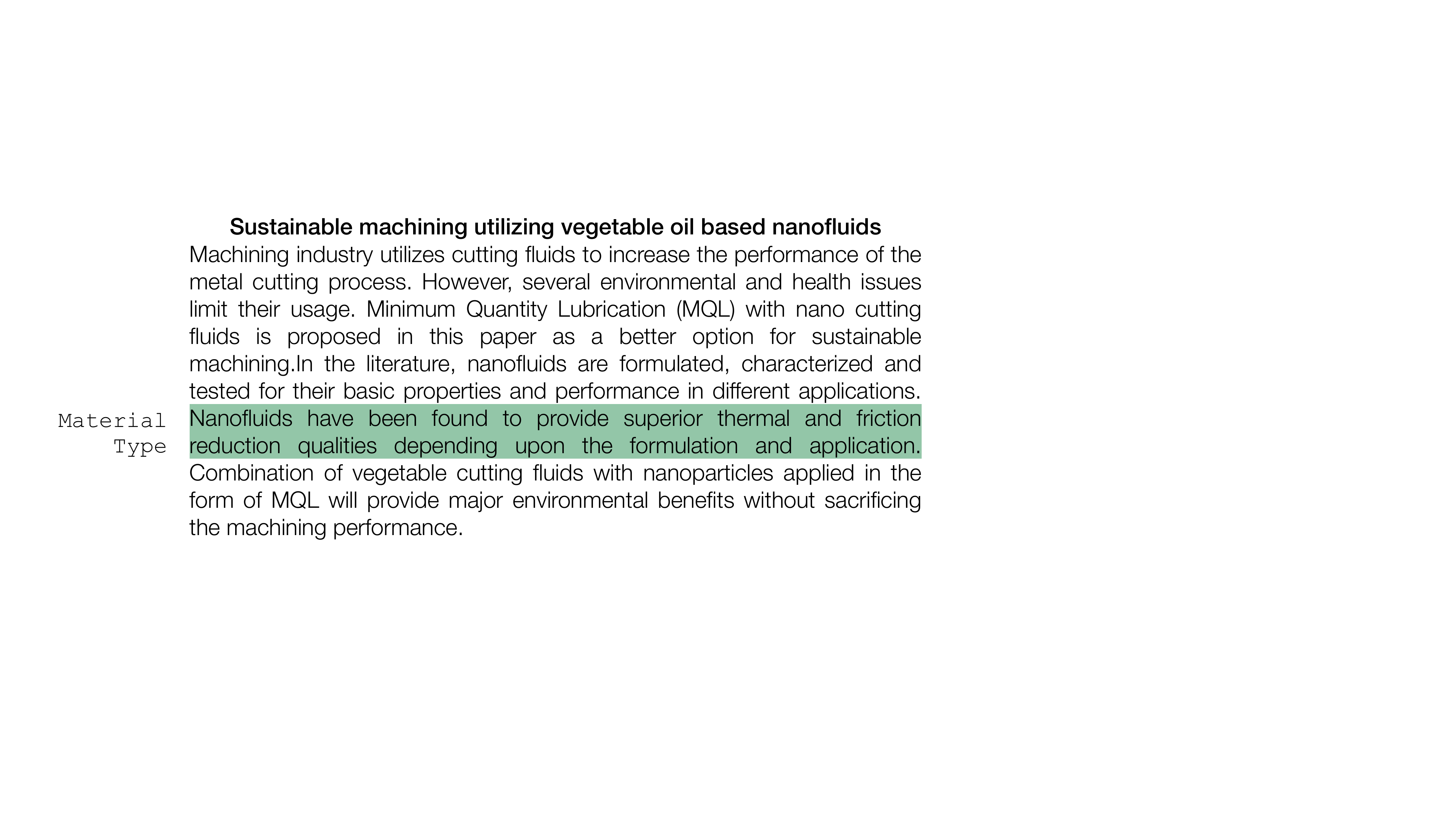}
    \vspace{-.75em}
    \caption{We leverage free-form sentence level querying \cirnum{\scriptsize 1} to retrieve documents across automatically built domain clusters \cirnum{\scriptsize 2}. These are expanded based on searcher selection and re-clustered to allow navigation of nuanced differences within selected global clusters \cirnum{\scriptsize 3}. Searchers can move between global clusters, `zoom in' to examine global clusters, and reformulate their queries for additional alternative exploration.}
    \label{fig:system-arch}
    \vspace{-1.5em}
\end{figure}

\begin{figure*}[t]
    \centering
    \includegraphics[width=.9\textwidth]{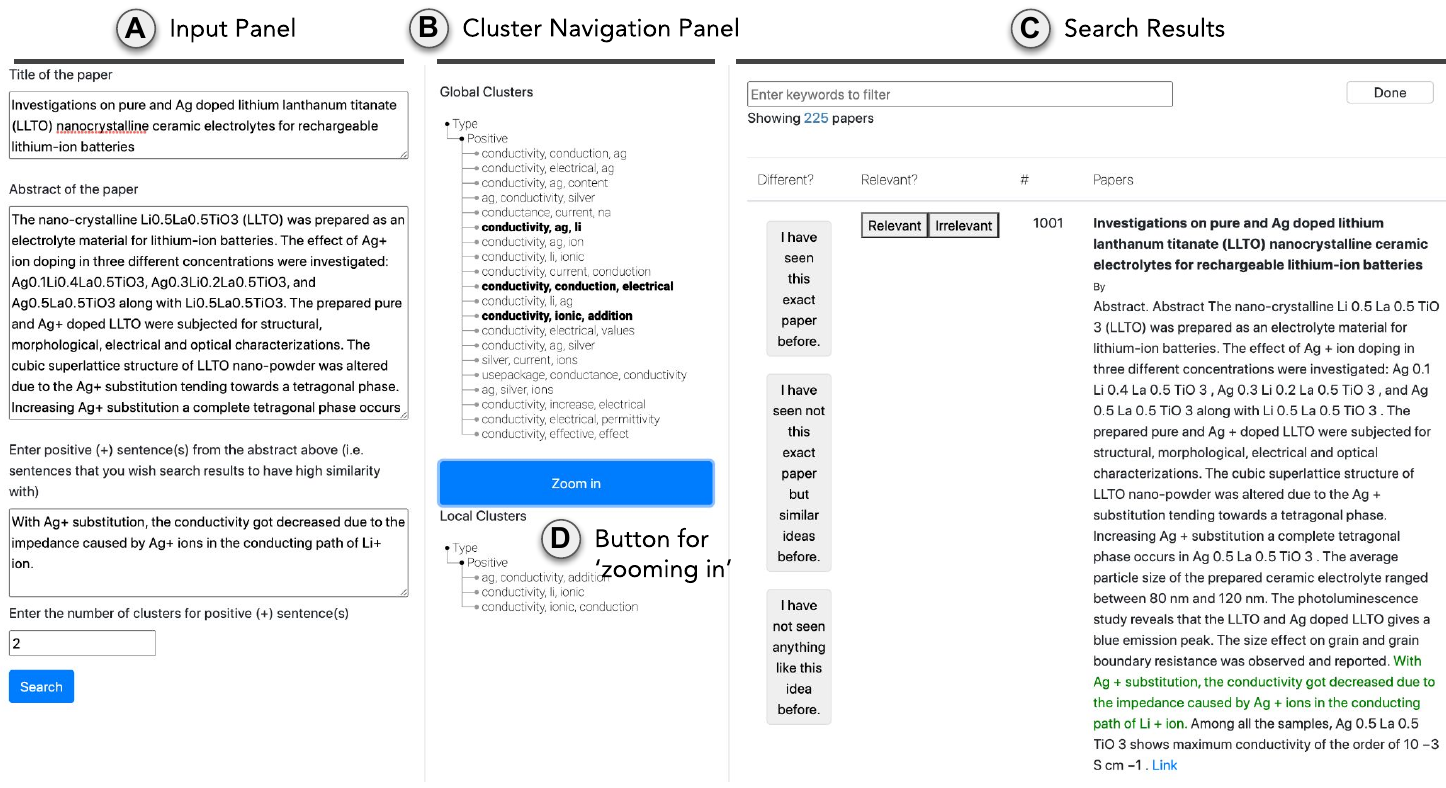}
    \vspace{-1em}
    \caption{Front-end interface. \cirnum{\scriptsize A} In the input panel, searchers enter the query paper abstract and a specific sentence aspect they want to align the retrieval results on. \cirnum{\scriptsize B} Once searchers click on `Search' at the bottom left, the system retrieves papers along with global domain clusters and displays them at the top of the navigation panel. Searchers can switch between different clusters to examine the results. Clicking on a cluster label highlights it and filters the papers that belong to the cluster in the table (clicking again on a selected cluster label de-selects it and refreshes the filtered table view). \cirnum{\scriptsize C} The papers are displayed in the table on the right, with a green sentence highlight showing the closest aspect to the query. Searchers can type keywords in the filter at the top to find papers that contain them (matching keywords in the abstract are highlighted). Optional control buttons for our case studies for collecting feedback data are also shown in the table. \cirnum{\scriptsize D} Clicking the `Zoom in' button retrieves more papers from the selected clusters, allowing searchers to examine nuanced differences within the selected domain clusters.}
    \vspace{-1.5em}
    \label{fig:interface}
\end{figure*}

Our system allows searchers to explore retrievals from global clusters and select interesting ones. The user may zoom into the clusters by retrieving additional results from them, which may reveal nuanced differences via local clustering. This interaction is supported by the `Zoom in' button in the interface (Fig.~\ref{fig:interface}). 
There are three main components of the system (Fig.~\ref{fig:system-arch}): \textit{Global clusters}. A set of clusters partitioning the entire corpus of papers into domains, built from paper embeddings computed with \citet{cohan2020specter}'s transformer-based model; \textit{Sentential representations}. Each sentence in an abstract is encoded using a state-of-the-art sentence encoder~\cite{MPNetBaseV2} to enable faceted retrieval at that level; \textit{Local clusters}. A set of nested clusters built from query-specific retrievals within user-specified global clusters to reveal variations within the clusters. For the text corpus of our system, we used 3.2M Materials Science abstracts in the S2ORC corpus~\cite{lo2020s2orc}. 

\xhdr{Global domain clusters} We represent broad sub-disciplines of materials science with a set of global clusters, $G\in\mathcal{G}$. These serve as the primary means for searchers to control the distance of retrievals to a query. The clusters were obtained via citation-based similarity clustering using \textsc{specter} embeddings of abstracts $d$'s in the corpus~\cite{cohan2020specter}. We employ a K-Means clustering of the embeddings with $|\mathcal{G}|=20$. This use follows from results indicating that pre-trained representations often capture domain structures well~\cite{aharoni2020unsupervised, peng2021neural}. We further validate these clusters in Appendix \ref{sec-cluster-eval}.

Following the creation of $\mathcal{G}$, we index the sentences $s$, of documents in each cluster, $d_g \in G$, with a cluster specific approximate nearest neighbour search index, $I_G$, in preparation for search. We use the HNSW index of Faiss \cite{johnson2019billion} and a pre-trained sentence encoder, $\mathtt{BERT}_s$ to encode the sentences \cite{MPNetBaseV2}.

\xhdr{Sentence-level retrieval} We leverage the recently proposed faceted QbE search paradigm~\cite{MysoreCSFCUBE2021} for querying, shown to support exploratory search tasks well~\cite{lissandrini2019exampletutorial}. Here, searchers query the system with an abstract $q$ and a sentence in the abstract $s_q\in q$ indicative of the aspect of their interest.

We embed $s_q$ with $\mathtt{BERT}_s$, and retrieve the $T$ nearest sentences from each global cluster index $I_G$. Users can then browse the results by switching between domains in the front-end interface, and additionally indicate specific global clusters $\mathcal{G}_s \subset \mathcal{G}$ they want to explore further. These serve as input to constructing query-specific local clusters.

\xhdr{Zooming in on global clusters with local clustering} Given a set of global clusters $\mathcal{G}_s$, and the query $s_q$, we retrieve the $L$ nearest sentences to $s_q$ from selected indices $I_{G\in\mathcal{G}_s}$. To support retrieval of more results from the same domain(s), we set $L > T$. We then re-cluster the similar document sentences into a set of $M$ clusters to show variations within the selected global clusters. 

Finally, our prototype used a preliminary \textsc{tfidf} scoring of unigrams in a cluster to extract descriptors for global and local clusters -- the informativeness of the descriptors was however limited.

\xhdr{Interface used in the case studies} \label{sec:interface} The main components of the front-end interface of our system are (See Fig.~\ref{fig:interface}) \circled{\small A}: Query input panel, \circled{\small B}: Cluster navigation panel, \circled{\small C}: Retrieval table view with filtering, and \circled{\small D}: An optional `Zoom in' button. Clicking this button retrieves more papers from the domain clusters searchers selected which allows them to explore and make sense of nuanced details among the results within the selected clusters. 

%% file: src2_study.tex
\vspace{-.5em} \section{Case Studies} \vspace{-.5em} We conducted think-aloud interviews~\cite{thinkaloud1} with three participants who engage in Materials Science and Engineering research to probe 1) what potentially \textit{distinct} values our system may bring to scientists exploring their own research questions compared to baseline systems, and 2) design challenges for realizing the full potential of search systems aimed at cross-domain exploration. We detail the study procedure in Appendix~\ref{appendix:study_procedure}.

%% file: src3_case_study_findings.tex
\vspace{-.5em} \section{Findings} \vspace{-.5em} \xhdr{Diverse results \& complementary value}
All participants commented on seeing more diverse results using our system over the baselines (Table~\ref{table:research_domains}). P1 compared his experiences as follows:
\begin{quoting}
``\textit{On ConnectedPapers (P1's baseline system), I can see the overall relations between works such as who's citing who, and identifying what's well-cited...but focused in the closely related fields like geochemistry and dissolution of cement...%
In [our system], you can see that \textbf{it's picking up things that I quite frankly had no idea why it picked up at first.}}'' -- P1
\end{quoting}
This quote demonstrates that while the prototype system did seem to retrieve more diverse results compared to the baseline system, the difference in results may require a closer look for apt engagement. Participants thought the value of retrieving such diverse results in the prototype system was complementary to that of the baseline systems. Both P2 and P3, who chose keyword query-based search engines as baselines, felt the baseline systems were useful for finding things in the domain they already knew: ``\textit{In regular search engines like \textsc{SciFinder} I can put in the exact query to find \textbf{precisely that kind of papers}}'' (P2). In comparison, ``\textit{[our system] is fantastic for learning more about a topic and kind of exploring...there are a lot of things that I found that I hadn't seen anywhere before.}'' (P2). However, this value also seemed to depend on how focused the search goal was: ``\textit{It depends on what I'm trying to do...If I had a specific experiment in mind, for example tracking the activity of magnesium...I probably wouldn't be interested in looking at even iron or non-manganese materials...but in other times, I'll be open to seeing even non-battery domains so long as similar challenges are addressed.}'' (P3).

\xhdr{Sourcing diversity from global domain clusters} This diversity seemed to originate from the differences in domains that participants perceived while interacting with the system. Specifically, participants thought the global clusters mapped to different application domains of similar mechanisms captured in their query sentences. For P1, this included clusters of papers focused on `mixture strategies to create cement with radiation shielding to use in military bases' (as opposed to recapturing materials from industrial wastes to create mixtures for buildings for non-military use) or `mixtures for bone-like scaffolds and implants in mice' (as opposed to buildings). In addition, P1 perceived the application domain of military bases as closer than Orthopedics in mice. An example P2 found interesting was a paper describing solid-state electrolytes for use in electrochromic mirrors, as opposed to the domain of battery technologies he was familiar with. This and other papers in the cluster showed P2 a whole new set of application domains related to optical properties of electrolytes that P2 did not know of. P3 found the results describing `discharging' mechanisms interesting and useful for learning new contexts of research that explore similar mechanisms, e.g., papers on electrostatic discharging observed in clouds or the effects of mechanical properties in anode coatings on discharging.

\xhdr{Zooming in \& realizing hidden assumptions} While reviewing the variations among the results, participants realized hidden assumptions that guided them to seek inspirations in `typical' domains. For example, in seeing varying uses of a similar mechanism (e.g., discharging) in different contexts (e.g., electrostatic vs. electrochemical%
), P3 learned about other domains which she did not tap into for inspirations before: ``\textit{Now I can see that discharging happens everywhere, even outside the battery domains. What I assumed was that discharging is only related to rate-related challenges in (an) electrochemical (context).}'' -- P3. Unlike similarity-maximizing search, learning about various domains helped P3 realize her own pre-conceived notion of typicality of a search domain, broaden potential domains of inspiration, and connect previously disjoint clusters (e.g., [mechanical strategies], such as nanoindentations in anode coating materials, to [electrochemical] discharging efficiency).

Retrieving more results in selected domain clusters and surfacing local clusters not only assisted participants in making sense of the high-level similarity within the clusters but also helped them see nuanced differences. For P2, seeing results on [doping] and [sintering temperature] were on-point with regards to the core mechanistic relevance for his experiments (high-level similarity), while 
zooming in the clusters revealed more nuanced trade-off relationships such as how doping has been used as a mitigation strategy against loss of lithium at high temperatures. Examining the additional retrievals also led P2 to notice diverse sub-fields such as [powder metallurgy] and [laser deposition] that the nuanced mechanistic differences occurred.

Sometimes recognizing such nuanced differences helped participants make sense of how the search system worked: ``\textit{So these papers lack (connection to) {dissolution} and are mostly about production of cement which makes sense because my query sentence only mentions producing aggregates. But I thought because the rest of the abstract talks about dissolution of aggregates, the system would know that's the context I'm focusing on.}'' -- P1. Interpreting how the system worked guided participants' query reformulation to counter their hidden assumptions or broaden the scope of search.

%% file: src3_implications.tex
\vspace{-.5em} \section{Implications for Design} \vspace{-.5em}
These results confirm earlier findings that emphasize support for an iterative discovery of important and generative \textit{mis-}alignments during analogical search~\cite{kang2022augmenting}, while also contributing new challenges specific to cross-domain exploration. Compared to similarity-maximizing search that optimizes for alignment of search results, cross-domain inspiration requires identification of aspects in analogs that may be diversified for end-user inspiration versus those that need to be preserved for retaining relevance.

Specifically, how to identify clusters that lack important connections to query problems is challenging because a cluster can be centered around a specific mechanism relevant to the query while lacking connections to other contextual mechanisms (e.g., a cluster of papers that focuses on production mechanisms of aggregate mixture but does not study the dissolution process of waste materials, which is a related mechanism for providing the ingredients to the production pipeline), or may not preserve important directionality of the cause-and-effect relationships (e.g., papers that demonstrate techniques to make electrolytes more inflammable may be useful in certain application domains but not for making battery technologies safer).

Given the importance of participants' reflection on their own assumptions and how the system performed search, future designs may focus on supporting \textit{stateful} explorations and aim to help searchers keep track of which clusters they have visited, what they found as useful (and not useful) in them, and how they might reformulate subsequent search queries based on the previously hidden but salient aspects of their search intent. This information may provide valuable feedback to the system for tuning its behavior over time.

Lastly, unfamiliar domains may require explanations of relevance for searchers to scaffold engagement at a deeper level and to prevent early rejection based on snap judgment. We observed that organizing papers in domain clusters based on their high-level similarity to the query aspect serves as an important anchor for sensemaking. Furthermore, encouraging end-users to zoom into local clusters that exhibit nuanced differences may have led our expert users to \textit{slow down} their judgment~\cite{carol2007slowing}, allowing a deeper processing of results. However, open questions remain for future work to effectively communicate papers' relevance to scientists (e.g.,~\citet{kang2022you}).

%% file: src4_conclusion.tex
\section{Conclusion} Our work presents an novel approach for exploring domain clusters to broaden sources of scientific inspiration and probe design implications for future interactive systems. Our case studies suggest that systems leveraging the global structure of knowledge domains and enabling flexible aspect-based retrieval help scientists find meaningful inspirations outside the domains they are familiar with. Additional work is needed to realize the full potential of interactive systems that augment scientific creativity via cross-domain inspirations.

\section{Acknowledgments}

We thank our study participants for their valuable insights and feedback. We would like to acknowledge partial funding from Shell, the Center for Knowledge Acceleration, the Allen Institute for Artificial Intelligence (AI2), National Science Foundation under Grant Number IIS-1763618, FW-HTF-RL-1928631, and IIS-1816242, DMREF Awards 1922311, 1922372, and 1922090, the Office of Naval Research (ONR) under contracts: N00014-20-1-2280 and N00014-19-1-2114, the IBM Research AI through the AI Horizons Network, and the Chan Zuckerberg Initiative under the project Scientific Knowledge Base Construction. This work is also supported by the Google Cloud Research Credits program with the award GCP19980904.

%% file: src5_appendix.tex
\section{Evaluation of Global Clusters}
\label{sec-cluster-eval}
To investigate the ability of clustering methods for separation of  materials science corpora into coherent subdomains, we compare model generated clusters with author-provided keywords treated as gold standard labels. 

\xhdr{Dataset} To generate our evaluation dataset domain experts first screened most frequent keywords in a multi-domain dataset of 700k English materials science publications and selected 18 keywords indicative of sub-domains (see Table \ref{table:cluster_eval-kw}). Next, only a subset of the corpus with documents assigned a single keyword of the 18 sub-domain keywords was retained for evaluation. This yields a corpus of 39,699 documents. A sample of these documents were checked manually to ensure that they represented reasonable paper-keyword assignments. This corpus is available.\footnote{\url{https://github.com/olivettigroup/cross-domain-exploration}}

\xhdr{Experimental Protocol} We compare the method used in our system, \textsc{specter}~\cite{cohan2020specter} to \textsc{tfidf} and \textsc{scibert} baselines \cite{beltagy2019scibert}. For evaluation of document clusters, embeddings are computed using publications titles and abstracts.  \textsc{scibert} and \textsc{specter} input the papers as "Title [SEP] Abstract". To generate clusters we employ K-Means with $K=18$. We report average cluster purity over three separate runs of clustering \cite[C16.3]{manningraghavan2008}. 

\xhdr{Results} Results (Table~\ref{table:cluster_eval}) show \textsc{specter}s' ability to induce subdomains clusters. Surprisingly high dimensional \textsc{tfidf} embeddings perform similarly to transformer generated \textsc{specter} embeddings (column: ``Keywords Present''). An explanation for this might be found in the evaluation dataset we created -- clusters based on author provided keywords, which are also likely to appear in the title and abstract of a paper are likely to see term-based methods (i.e.\ \textsc{tfidf}) seeing strong performance. To investigate this further, we repeat the evaluation  after removing the 18 category defining keywords (Table \ref{table:cluster_eval-kw}) from abstracts and title text before encoding papers. This experiment (column: ``Keywords Removed'') strengthens our hypothesis, showing a larger drop in the measured cluster purity values for the \textsc{tfidf} approach compared to \textsc{specter}. \textsc{specter}'s ability to induce clusters in the absence of explicit term overlap provides the value needed in our search setup.

\begin{table}[t!]
\centering
\begin{tabular}{l>{\centering\arraybackslash}p{.1\textwidth}>{\centering\arraybackslash}p{.1\textwidth}}\toprule
        & Keywords Present & Keywords Removed \\\midrule
\textsc{random}  & 27.37         & 27.37            \\
\textsc{tfidf}   & \textbf{48.08}         & 39.75            \\
\textsc{scibert} & 31.80         & 31.32            \\
\textsc{specter} & 47.93         & \textbf{43.90}    \\\bottomrule       
\end{tabular}
\caption{Domain cluster purity for \textsc{specter}, compared to baseline approaches on a corpus of materials science papers and author-provided keywords. Number of clusters = 18. Reported metrics represent cluster purity in percent, averaged over three runs. ``Keywords Removed'' represents the performance after removal of all category defining keywords (of Table \ref{table:cluster_eval-kw}) from the text of the title and abstract based on which paper representations are computed.}
    \label{table:cluster_eval}
\end{table}

\begin{table}[t!]
\center
\begin{tabular}{cc}
\toprule
magnetic materials & carbon materials \\
ceramics & optical properties\\
electrochemistry & nanomaterials\\
alloys & photocatalysis\\
semiconductors & solar cells\\
fuel cells & polymers\\
composite materials & biomaterials\\
thermodynamics & lithium-ion batteries\\
thin films & microstructure\\
\bottomrule
\end{tabular}
\caption{Keywords representative of clusters in our cluster evaluation dataset.}
    \label{table:cluster_eval-kw}
\end{table}

\section{Procedure and Interface Design Used in the Case Studies} \label{appendix:study_procedure}
\xhdr{Participants} Three participants (1 female) whose research areas are in the broader space of material science and engineering were recruited for interviews. The mean age of participants was 24 (SD=4.04). They all reported conducting a review of related work using scholarly search engines as a research facilitator and vehicle for learning.

\begin{table*}
    \centering
    \begin{tabular}{P{.04\textwidth} p{.45\textwidth} P{.2\textwidth} P{.21\textwidth}}
    \toprule
    \multirow{2}{*}{\textbf{PID}} & \multirow{2}{*}{\textbf{Research Domains}} & \textbf{Seed Paper \& Query sentence} & \textbf{Baseline Search System} \\
    \midrule
    \multirow{3}{*}{1} & New ways of producing cement using industrial wastes to lower the carbon footprint and maintain sustainability in commercial bldgs & \small{Query sentence was the 2nd sentence in the abstract of~\citet{seed_p1}} & \multirow{3}{*}{\textsc{ConnectedPapers}\footnote{\url{https://www.connectedpapers.com/}}} \\
    \multirow{2}{*}{2} & New electrolyte materials for simultaneously increasing the capacity and stability of batteries & \small{Query sentence: 8th abstract sentence in~\citet{seed_p2}} & \multirow{2}{*}{\textsc{SciFinder}\footnote{\url{https://scifinder.cas.org/scifinder}}} \\
    \multirow{2}{*}{3} & Understanding the barriers to scaling up the manufacturing of effective cathode materials & \small{Query sentence: 2nd abstrct sentence in~\citet{seed_p3}} & \multirow{2}{*}{\textsc{Google Scholar}}\\
    \bottomrule
    \end{tabular}
    \caption{Participants' research domains, seed papers used in the study, and baseline search engines they chose.}
    \label{table:research_domains}
    \vspace{-1em}
\end{table*}

\xhdr{Procedure} Participants prepared a query paper and a description of their research problem (Table~\ref{table:research_domains}) and chose search engines they personally use the most and are familiar with for the first stage (20 min) of the interview. Using the search engines of their choice, participants in the first stage looked for as many related work as possible for the query domain. The goal of this stage was to gain a sense of `what is out there' which served as context for their judgement on retrieval results in the subsequent stage using our system. In the next stage, participants interacted with our system with the focus of finding different ideas than what they have found in the first stage (20 min). For the papers they found interesting, they provided feedback on two dimensions, 1) \textit{Novelty} with the following ternary response options (1: ``\textit{I have seen this exact paper before}''; 2: ``\textit{I have not seen this exact paper but similar ideas before}''; 3: ``\textit{I have not seen anything like this before},'' similar to the scale used in~\cite{kang2022augmenting}) and 2) \textit{Relevance} with binary response options. The collected feedback was used in the end-of-session debriefs to aid both the interviewer's and participants' memory and discussion grounded on specific examples.

\section{Related Work}
\label{sec-extended-related-work}
Our work bears resemblance to a number of prior lines of work. We group these broadly into methods for exploratory search of text corpora, and methodological NLP and information retrieval work which bears resemblance to parts of our system.

\subsection{Exploratory Search for Text Corpora}
A large body of work has explored the problem of exploratory search for text corpora, with searchers often interested in learning and discovery of new content. These search tasks are often characterized by an open-ended and multifaceted information need and involve a sizable browsing component \cite[Sec 2.4]{White2009ExploratorySB}. Given the importance of these aspects, we group the ways in which prior work represents documents in a system, this in turn, affect the ways that users query a search system. In describing past work we also highlight the ways in which they facilitate browsing.

\xhdr{Exploration via rhetorical structure} This line of work primarily consists of designing a schema of labels to describe the rhetorical structure elements of documents, automatically or manually extracting these elements from documents, and indexing and retrieving documents along these aspects in response to a query. Document representation with rhetorical structure elements follows from a large body of work demonstrating that problem solving is often characterized by descriptions of the Situation, Problem, Solution, and Evaluation \cite[Sec 1]{heffernan2018identifying}. In practice, a primary bottleneck of this line of work is its reliance on a schema of labels - making transitions to new corpora challenging. 
The closest prior work of this type is that of analogical search for creative ideation applied to science and design \cite{hope_kdd17, hope2021scaling, kang2022augmenting}. The method of query specification in this body of work is primarily via short text queries. In contrast, our system adopts a faceted query-by-example approach where searchers query with an example document and indicate an aspect of the document they would to focus on via selecting text spans, with documents indexed at the sentence level \cite{MysoreCSFCUBE2021}. 
In this respect, work of \citet{chan2018solvent} and \citet{neves2019evaluation} bear the closest resemblance to ours in supporting querying with documents in combination with a fixed schema of aspect labels. 

While most work highlighted above retrieve a list of documents, \citet{chan2018solvent} and \citet{hope2021scaling} present exceptions in allowing browsing via domain-like structures - bearing resemblance to our approach. While \citet{chan2018solvent} search across domain structures informed by document metadata, our system relies on a set of automatically built global domain clusters and a set of query specific local clusters. Similarly, \citet{hope2021scaling}, allow exploration of purpose and mechanism aspects via a hierarchical organization of these aspects constructed with a rule-mining approach. This approach, while well suited to use cases where aspects can be pre-extracted proves challenging to operationalize in the absence of these extractions. Our work complementarily prioritizes flexibility in query specification without a schema based document representation.

Other work also bear resemblance to the ones presented above. \citet{jain2018learning} learn aspect disentangled paper representations with \emph{aspect specific} encoders trained with similarity data aligned to aspects in biomedical abstracts. \citet{ferosa2016tanmoy} present a system for retrieval of similar papers grouped along a schema of aspects based on citation and content similarities to a query. \citet{ostendorff2020aspectcoling}, present an approach to classify the aspect of similarity between a pair of papers rather than producing a ranking over documents. \citet{portenoy2022bridger} present a system for exploring \textit{authors} across knowledge domains.

\xhdr{Exploration via topical structures} A separate line of work has also leveraged topical structures such as knowledge base entities (concepts), keywords, or other model inferred topic structures. \citet{hope2020scisight} present a system for exploration of biomedical concepts (e.g. proteins, diseases, drugs) and research groups. Similarly, \citet{shukla2021visually} enhance academic search engine result pages with keywords available in document metadata, and \citet{sorkhei2017exploring} allow exploration of a corpus of papers via topics inferred with a LDA model. Besides the bottleneck of concept extraction in some of these methods, a key challenge of concept based exploration, as noted in \citet{hope2020scisight} and \citet{sultanum2020understanding}, stems from users desire to continually explore additional concepts unavailable in a system or add concepts of their own. We alleviate this drawback in part by allowing users to directly query with text spans in a document context. That said, the two methods provide complementary value -- while topical structures provide value in gaining an overview of a corpus our approach is likely to allow expression of more complex queries.

\subsection{Search and Recommendation}
A body of work in NLP, information retrieval, and recommendation bear resemblance to parts of our system. We group these into query-by-example methods and methods for cluster ranking and result diversification. While the former contextualizes our document representation and querying mechanism the latter contextualizes our result presentation. 

\xhdr{Query-by-Example} This body of work primarily intends to retrieve documents given a query document and is well suited to exploratory search given its ability to express more context for under-specified queries \cite{ksikes2014towards, lissandrini2019exampletutorial}. This is best represented in \citet{el2011qbepapers}, who present a system for personalized recommendation of papers given a set of user trusted papers - with documents presented as a diversified ranking of candidate documents based on citation network similarities. A large body of work under citation recommendation also recommends documents similar to a query document \citet{farber2020citation}. Recent work has also seen strong performance for a variety of scientific paper recommendation tasks by training transformer bi-encoder models on citation network data \cite{cohan2020specter, Ostendorff2022SciNCL}. In contrast with this work, our work intends to model aspect conditional similarities. Closest work to ours is presented by \citet{mysore2021aspire} who extend these transformer bi-encoder models to a sentence level multi-vector model intended for finer-grained aspect conditional document similarities trained with co-citation data and textual supervision.

\xhdr{Cluster retrieval and diversification} In our generation of results based on a set of global clusters and a set of local clusters, our work departs from the standard IR assumption of treating candidate documents as independent from each other - \citet{Pang2021BeyondPRP} overview this line of work. In doing so, our work relates to lines of work on search result diversification and cluster retrieval. Work in search result diversification most often intends to improve retrieval performance for ambiguous queries by optimizing the novelty of retrieved results or optimizing for coverage of all possible interpretations of a query, \citet{santos2015diversesurvey} surveys this body of work in the context of search, while \citet{kaminskas2016diversity} survey similar work in the context of recommendations. Close work in cluster retrieval comes from \citet{Levi2018selectivecluster}, who investigate the appropriate queries for which results should be grouped on a search result page. These lines of work differ from ours in intending to improve performance for ad-hoc search, our work on the other hand intends to use clusters of documents as a means of controlling ``distance'' to a query in an exploratory search setup. In this respect, \citet{Xie2021domaincite} represents closely related recent work in training a model for citation recommendation across-domain boundaries while leveraging an existing ontology of domain concepts to learn the domain specific semantics - training similar models for our setup represent exciting future steps. 

Finally, cross-domain retrieval \cite{cohen2018cross, tran2019domain} aims to improve in-domain performance of retrieval models in a transfer learning setup while our work focuses on retrieval across domain boundaries.